\documentclass[aps,superscriptaddress,twocolumn,a4paper]{revtex4}
\usepackage[colorlinks=true, pdfstartview=FitV, linkcolor=blue, citecolor=red, urlcolor=black]{hyperref}
\usepackage{graphicx}
\usepackage[all]{xy}
\usepackage{amsmath}
\usepackage{amssymb}
\usepackage{tensor}
\newcommand{\be}{\begin{equation}}
\newcommand{\ee}{\end{equation}}
\newcommand{\ben}{\begin{eqnarray}}
\newcommand{\een}{\end{eqnarray}}
\newcommand{\bes}{\begin{subequations}}
\newcommand{\ees}{\end{subequations}}
\def\bal#1\eal{\begin{align}#1\end{align}}
\newcommand{\ov}{\overline}

\newcommand{\nn}{\nonumber\\}
\newcommand{\bfi}{\begin{figure}}
\newcommand{\efi}{\end{figure}}
\newcommand{\bc}{\begin{center}}
\newcommand{\ec}{\end{center}}

\newcommand{\LL}{{\cal L}}
\newcommand{\Sc}{{\cal S}}

\newcommand{\p}{\partial}

\newcommand{\vphi}{\varphi}
\newcommand{\vphic}{\ov{\varphi}}
\newcommand{\vphia}{|\varphi|}

\begin{document}

\title{First order framework
for vortices in generalized \\ Maxwell-Chern-Simons models
without a neutral field}
\author{I. Andrade}\affiliation{Departamento de F\'\i sica, Universidade Federal da Para\'\i ba, 58051-970 Jo\~ao Pessoa, PB, Brazil}
\author{D. Bazeia}\affiliation{Departamento de F\'\i sica, Universidade Federal da Para\'\i ba, 58051-970 Jo\~ao Pessoa, PB, Brazil}
\author{M. A. Liao}\affiliation{Departamento de F\'\i sica, Universidade Federal da Para\'\i ba, 58051-970 Jo\~ao Pessoa, PB, Brazil}
\author{M. A. Marques}\affiliation{Departamento de Biotecnologia, Universidade Federal da Para\'\i ba, 58051-900 Jo\~ao Pessoa, PB, Brazil}\affiliation{Departamento de F\'\i sica, Universidade Federal da Para\'\i ba, 58051-970 Jo\~ao Pessoa, PB, Brazil}
\author{R. Menezes}\affiliation{Departamento de Ci\^encias Exatas, Universidade Federal
da Para\'{\i}ba, 58297-000 Rio Tinto, PB, Brazil}
\affiliation{Departamento de F\'\i sica, Universidade Federal da Para\'\i ba, 58051-970 Jo\~ao Pessoa, PB, Brazil}


\begin{abstract}
This work introduces a procedure to obtain vortex configurations described by first order equations in generalized Maxwell-Chern-Simons models without the inclusion of a neutral field. The results show that the novel methodology is capable of inducing important modification in the vortex core, leading to vortex configurations with unconventional features.
\end{abstract} 

\maketitle

\section{Introduction}
Vortices are planar structures that arise in High Energy Physics under the action of a complex scalar field minimally coupled to a gauge field via an Abelian $U(1)$ local symmetry. The first relativistic model supporting vortex solutions was proposed in Ref.~\cite{NO}, with the gauge field guided by the standard Maxwell dynamics. The vortex configurations are electrically neutral but support magnetic flux that is quantized by the vorticity, which is associated to the topological character of the structure. These objects were also investigated in models whose gauge field obeys the Chern-Simons dynamics \cite{jackiw,coreanos}. In this situation, the vortex solutions engender both electric charge and magnetic flux quantized by the vorticity. In Refs.~\cite{paulkhare,jacobs}, vortices were investigated in models with both Maxwell and Chern-Simons terms.

The study of vortices, regardless the scenario considered, involves equations of motion that are of second order. They usually present couplings between the functions that describe the structure and this makes the problem hard to solve. To obtain first order equations, one may consider stressless solutions, which are stable under contractions and dilations \cite{derrick,bogopaper,schaposnik,godvortex}, and appear for the potencial written in a specific manner. In the case of Maxwell or Chern-Simons, this can be done without the presence of extra fields. Nevertheless, to perform this task in the Maxwell-Chern-Simons (MCS) scenario, one has to include a neutral field to balance the component $A_0$ of the gauge field \cite{nmcs,bazeiamcs}. Even in noncanonical models, the neutral field is required to obtain first order equations in the MCS case \cite{godvortex,menezesmcs}. A manner to throw the neutral field away is by considering a nonminimal coupling between the fields \cite{torres,ghoshplb,ghosh,nminimo}. However, the challenge here is to maintain the fields minimally coupled, without the presence of neutral field.   

The investigation is inspired by the recent Refs.  \cite{godvortex,nminimo}, with the aim of introducing a procedure to get vortex solutions described by first order equations in a generalized MCS model that excludes neutral field and nonminimal coupling between the matter and gauge fields. To implement the study, we organize the work as follows. In Sec. \ref{sec2} we describe the general model and obtain the first order equations using two distinct procedures, one described in \cite{godvortex}, and the other in \cite{bogopaper}. Moreover, in Sec. \ref{sec3} we illustrate our results investigating two distinct models, giving rive to vortices with distinct internal structures. We end the work in Sec. \ref{end}, adding comments on other related lines of investigation.

\section{The general model}\label{sec2}

Considering the metric tensor $\eta_{\alpha\beta}=\text{diag}(+,-,-)$, we work with the action in the form $\Sc=\int d^3x\,\LL$, where the Lagrange density includes both Maxwell and Chern-Simons terms. The generalized model to be studied in the present work includes two distinct quantities: $\mu(\vphia)$, which is a generalized magnetic permeability, and $M(\vphia)$, which modifies the dynamics of the scalar field. The Lagrange density of the model has the form
\be\label{model}
\begin{aligned}
\LL &= -\frac{1}{4\mu(\vphia)}F_{\alpha\beta}F^{\alpha\beta} +\frac{\kappa}{4}\epsilon^{\gamma\alpha\beta}A_\gamma F_{\alpha\beta}\\
	&+M(\vphia)\ov{D_\alpha\vphi}D^\alpha\vphi - V(\vphia).
\end{aligned}
\ee
In the above expression, $F_{\alpha\beta}=\p_\alpha A_\beta - \p_\beta A_\alpha$ is the electromagnetic strength tensor and we consider the scalar field minimally coupled to the gauge field, with $D_\alpha\vphi=\p_\alpha\vphi +ieA_\alpha\vphi$. We also consider natural units $(\hbar=c=1)$. By doing so, the dimensions of the several quantities are, using $\xi$ to represent the dimension of energy: $[x^\alpha]=\xi^{-1}$, $[\vphi]=[A^\alpha]=[e]=\xi^{\frac12}$, $[\kappa]=\xi^1$, and $[V(\vphia)]=\xi^{3}$. Moreover, both $\mu(\vphia)$ and $M(\vphia)$ are dimensionless functions of $\vphia$. 

The above model \eqref{model} gives the equations of motion for the scalar and gauge fields
\bes\label{eom}
\bal\label{eomphi}
&D_\alpha\left(M(\vphia)D^\alpha\vphi\right) = \frac{\vphi}{2\vphia}\bigg(\frac{\mu_{\vphia}}{4\mu^2(\vphia)}F_{\alpha\beta}F^{\alpha\beta}\nn
	&+M_{\vphia}\ov{D_\alpha\vphi}D^\alpha\vphi -V_{\vphia}\bigg),\\
\label{eomA}&\p_\alpha\left(\frac{F^{\alpha\gamma}}{\mu(\vphia)}\right) +\frac{\kappa}{2}\epsilon^{\gamma\alpha\beta}F_{\alpha\beta} = J^\gamma.
\eal
\ees
Here, $J_\alpha=ieM(\vphia)(\vphic D_\alpha\vphi -\vphi\ov{D_\alpha\vphi})$ is the current and we use the notation $\mu_{\vphia}=\p\mu/\p\vphia$ and so on. The energy-momentum tensor associated to the Lagrange density \eqref{model} reads
\be\label{tmunu}
\begin{aligned}
T_{\alpha\beta} &= \frac{1}{\mu(\vphia)}\left(F_{\alpha\gamma}\tensor{F}{^\gamma_\beta} +\frac14\eta_{\alpha\beta}F_{\gamma\sigma}F^{\gamma\sigma}\right) +\eta_{\alpha\beta}V(\vphia)\\
	&+M(\vphia)\left(\ov{D_\alpha\vphi}D_\beta\vphi +D_\alpha\vphi\ov{D_\beta\vphi} -\eta_{\alpha\beta}\ov{D_\gamma\vphi}D^\gamma\vphi\right).
\end{aligned}
\ee

To deal with vortex configurations, we consider static fields with the usual ansatz
\bes\label{ansatz}
\bal
&\vphi = g(r)e^{in\theta},\\ 
&A_0 = h(r),\\
&\vec{A} = \frac{\hat{\theta}}{er}\left(n -a(r)\right),
\eal
\ees
where $n$ is an integer number that controls the vorticity (winding number) of the field configurations. The above functions must obey the boundary conditions
\bes\label{bcond}
\bal
&a(0) = n, & a(\infty) &\to 0,\\
&g(0) = 0, & g(\infty) &\to v,\\
\label{bcondh}&h(0) = h_0, & h(\infty) &\to h_\infty.
\eal
\ees
The electric and magnetic fields, defined as $E^i = F^{i0}$ and $B =-F^{12}$, have the following forms
\be\label{fieldsEB}
\vec{E} = -h^\prime\,\hat{r} \quad\text{and}\quad B = -\frac{a^\prime}{er},
\ee
where the prime denotes derivative with respect to spatial coordinate $r$. By integrating the above magnetic field in the $(r,\theta)$ plane, one gets the quantized flux $\Phi = 2\pi n/e$.

With the above field configurations, the equation of motion \eqref{eomphi} associated to the scalar field reads
\be\label{scalareom}
\begin{aligned}
&\frac{1}{r}\left(rMg^\prime\right)^\prime +gM\left(e^2h^2 -\frac{a^2}{r^2}\right) +\frac14\frac{\mu_g}{\mu^2}\left(\frac{{a^\prime}^2}{e^2r^2} -{h^\prime}^2\right)\\
&+\frac12M_g\left(e^2g^2h^2 -{g^\prime}^2 -\frac{g^2a^2}{r^2}\right) -\frac12V_g= 0,
\end{aligned}
\ee
and Gauss' and Amp\`ere's laws from Eq.~\eqref{eomA} become
\bes\label{mcseom}
\bal \label{gausslaw}
&\frac{1}{r}\left(\frac{rh^\prime}{\mu}\right)^\prime -\frac{\kappa a^\prime}{er} -2e^2g^2Mh = 0,\\ \label{amplaw}
&\left(\frac{a^\prime}{er\mu}\right)^\prime -\kappa h^\prime -\frac{2eg^2Ma}{r} = 0.
\eal
\ees
The energy density with the ansatz \eqref{ansatz} is obtained from $T_{00}$ in Eq.~\eqref{tmunu}, as
\be\label{rho}
\rho = \frac{1}{2\mu}\left({h^\prime}^2 +\frac{{a^\prime}^2}{e^2r^2}\right) +M\left(e^2g^2h^2 +{g^\prime}^2 +\frac{g^2a^2}{r^2}\right) +V.
\ee
The other components of the energy momentum tensor are
\bes
\bal
T_{01} &= -\frac{1}{r}\left(\frac{a^\prime h^\prime}{e\mu} +2eg^2Mah\right)\sin(\theta),\\
T_{02} &= \frac{1}{r}\left(\frac{a^\prime h^\prime}{e\mu} +2eg^2Mah\right)\cos(\theta),\\
T_{12} &= \left(M\left({g^\prime}^2 -\frac{a^2g^2}{r^2}\right) -\frac{1}{2\mu}{h^\prime}^2\right)\sin(2\theta),\\
T_{11} &= \frac{1}{2\mu}\frac{{a^\prime}^2}{e^2r^2} +e^2g^2Mh^2 -V\\
	&+\left(M\left({g^\prime}^2 -\frac{a^2g^2}{r^2}\right) -\frac{1}{2\mu}{h^\prime}^2\right)\cos(2\theta),\\ 
T_{22} &= \frac{1}{2\mu}\frac{{a^\prime}^2}{e^2r^2} +e^2g^2Mh^2 -V\\
	&+\left(M\left(\frac{a^2g^2}{r^2} -{g^\prime}^2\right) +\frac{1}{2\mu}{h^\prime}^2\right)\cos(2\theta).
\eal
\ees

The equations of motion \eqref{scalareom} and \eqref{mcseom} are of second order. To get first order equations, we follow the lines of Ref.~\cite{godvortex} and take the rescale $r\to\lambda r$ in the functions $a(r)$, $g(r)$ and $h(r)$. So, we take $a(r)\to a^{(\lambda)}=a(\lambda r)$, $g(r)\to g^{(\lambda)}=g(\lambda r)$ and $h(r)\to h^{(\lambda)}=h(\lambda r)$, and use Eqs.~\eqref{gausslaw} and \eqref{rho} to calculate the energy of the rescaled solutions, which is written as
\be
\begin{aligned}
E^{(\lambda)} &= 2\pi\int rdr\,\Bigg(\frac1r\frac{\p}{\p r}\left(\frac{1}{\mu}rh^{(\lambda)}\frac{\p h^{(\lambda)}}{\p r}\right)\\
	&-\frac{\kappa}{er}h^{(\lambda)}\frac{\p a^{(\lambda)}}{\p r} +\frac{1}{2\mu}\!\left(\!\frac{1}{e^2r^2}\left({\frac{\p a^{(\lambda)}}{\p r}}\right)^{\!\!2} -\left({\frac{\p h^{(\lambda)}}{\p r}}\right)^{\!\!2}\right)\\
	&+M\!\left(\!\left(\frac{\p g^{(\lambda)}}{\p r}\right)^{\!\!2} +\frac{{a^{(\lambda)}}^2{g^{(\lambda)}}^2}{r^2} -e^2{g^{(\lambda)}}^2{h^{(\lambda)}}^2\right)\! +V\Bigg)\\
	&= 2\pi\int zdz\lambda^{-2}\,\Bigg(\lambda^2\frac{1}{z}\frac{\p}{\p z}\left(\frac{1}{\mu}zh\frac{\p h}{\p z}\right)\\
	&-\lambda^2\frac{\kappa}{ez}h\frac{\p a}{\p z} +\frac{1}{2\mu}\left(\lambda^4\frac{1}{e^2z^2}\left({\frac{\p a}{\p z}}\right)^2 -\lambda^2\left(\frac{\p h}{\p z}\right)^2\right)\\
	&+M\left(\lambda^2\left(\frac{\p g}{\p z}\right)^2 +\lambda^2\frac{a^2g^2}{z^2} -e^2g^2h^2\right) +V\Bigg)
\end{aligned}
\ee
where $z=\lambda r$.  To get rotationally symmetric solutions, we take $T_{12}=0$, which leads us to
\be\label{der1}
M\left({g^\prime}^2 -\frac{g^2a^2}{r^2}\right) -\frac{1}{2\mu}{h^\prime}^2 = 0,
\ee
and also to $T_{11}=T_{22}$. Since the energy-momentum tensor is conserved, one can write $T_{11}=T_{22}=C$, where $C$ is constant. To ensure stability under rescaling, i.e., that solutions with $\lambda=1$ minimize the energy, we impose $\p E^{(\lambda)}/\p\lambda\big|_{\lambda=1}=0$ and $\p^2 E^{(\lambda)}/\p\lambda^2\big|_{\lambda=1}>0$. These conditions are satisfied by $T_{11}=T_{22}=0$, which reads
\be\label{der2}
e^2g^2Mh^2 +\frac{1}{2\mu}\frac{{a^\prime}^2}{e^2r^2} - V = 0.
\ee
This means that the stressless condition $T_{ij}=0$ leads to rotationally symmetric solutions stable under rescale. As shown in Ref.~\cite{godvortex}, first order equations for vortices in Maxwell-Chern-Simons models can be obtained with the addition of a neutral scalar field to balance the contribution of $h(r)$ in the stress, $T_{ij}$. Here, we introduce a novel procedure to obtain first order equations without the presence of the neutral field. We have found that this can be achieved if $\mu(\vphia)$ and $M(\vphia)$ are related by
\be\label{M}
M(\vphia) = \frac{\kappa^2\ell\left(\ell+1\right)\mu(\vphia)}{2e^2\vphia^2}.
\ee
where $\ell$ is a dimensionless real parameter introduced as above for convenience. Since both $\mu(\vphia)$ and $M(\vphia)$ are non-negative functions, we must have $\ell(\ell+1)>0$. Moreover, to avoid divergencies in $M(\vphia)$, we choose the magnetic permeability such that there is a real number $L$ for which
\be\label{lim}
\lim_{\vphia\to0}\frac{\mu(\vphia)}{\vphia^2}=L,
\ee
i.e., $\mu(\vphia)$ goes to zero at least as fast as $\vphia^2$. 

The issue now is to unveil the first order framework for the above model, in the absence of an extra neutral scalar field. After several unsuccessful attempts, inspired by Eq. \eqref{der1} we have found an interesting possibility. It is implemented with the equation
\be\label{hprime}
-h^\prime = \frac{\kappa\ell\mu a}{er}.
\ee
By using it in Gauss' law Eq.~\eqref{gausslaw}, we get
\be\label{bh}
-\frac{a^\prime}{er} = \kappa\ell\mu h.
\ee
From Eqs.~\eqref{M} and \eqref{bh}, one may show that the charge density has the form $
J_0 =\kappa\left(\ell+1\right)a^\prime/(er)$. By integrating it, one gets the electric charge $Q=-\kappa(\ell+1)\Phi$. Since the sign of magnetic flux is not influenced by $\ell$, we see that the electric charge may change its sign depending on $\ell$.

By using the latter three equations one can show that Amp\`ere's law \eqref{amplaw} is solved. By substituting them in Eqs.~\eqref{der1} and \eqref{der2}, one gets the first order equations
\bes\label{fo}
\bal\label{fog}
g^\prime &= \pm\sqrt{\frac{2\ell+1}{\ell+1}}\frac{ag}{r},\\ \label{foa}
-\frac{a^\prime}{er} &= \pm\sqrt{\frac{2\ell\mu(g) V(g)}{2\ell+1}}.
\eal
\ees
These equations must be solved according to the boundary conditions in Eq.~\eqref{bcond}. Notice that there is an algebraic factor guided by $\ell$ in the right hand side of Eq.~\eqref{fog}. This feature is new in the study of vortices in models with minimal coupling, at first order, where one finds that $g(r)$ satisfies $g^\prime=ag/r$, which leads to $g(r\approx0)\propto r^{|n|}$ in these systems; see Ref.~\cite{godvortex}. It resembles the first order equations associated to models with non-minimal coupling \cite{ghoshplb,ghosh,nminimo}.
The aforementioned factor modifies the behavior of $g(r)$ near the origin. One can show that $g(r)$ behaves as
\be\label{gori}
g(r\approx0)\propto r^{|n|\sqrt{(2\ell+1)/(\ell+1)}}.
\ee
The behavior of $a(r)$ depends on the model.

The first order equations \eqref{fo} determine the form of the functions $a(r)$ and $g(r)$. By knowing them, one can find $h(r)$ from Eqs.~\eqref{bh} and \eqref{foa}, as
\be\label{h}
h(r) = \pm\frac{1}{\kappa\ell}\sqrt{\frac{2\ell V(g)}{\left(2\ell+1\right)\mu(g)}}.
\ee
Thus, the solutions are completely determined by Eqs.~\eqref{fo} and \eqref{h}. In these equations, the upper/lower sign stands for positive/negative vorticity. The case of positive vorticity is related to the one with negative vorticity by the change $a\to-a$ and $h\to-h$.

To ensure that equations \eqref{fo} and \eqref{h} are compatible with the equations of motion \eqref{scalareom} and \eqref{mcseom}, we take the constraint
\be\label{vinc}
\frac{d}{dg}\sqrt{\frac{2V}{\mu}} = -\frac{\kappa^2\sqrt{\ell^3\left(\ell+1\right)}\mu}{eg},
\ee
which leads to the potential
\be\label{pot}
V(g) = \frac{\kappa^4\ell^3\left(\ell+1\right)\mu(g)}{2e^2}\left(\int dg\,\frac{\mu(g)}{g}\right)^2.
\ee
So, the potential is specified by the generalized magnetic permeability, $\mu(g)$, which must be chosen to allow for the existence of solutions obeying the boundary conditions in Eq.~\eqref{bcond}. To comply with Eq.~\eqref{lim}, we take magnetic permeabilities with $\mu(0)=0$ whose associated potential supports a minimum at $g=0$, with $V(0)=0$.

We can rewrite the energy density \eqref{rho} using Eqs.~\eqref{hprime} and \eqref{bh}, substituting $h$ and $h^\prime$ to make it to depend only on $g$ and $a$ and their corresponding derivatives. This gives
\be
\rho = \frac{(2\ell +1){a^\prime}^2}{2e^2\ell r^2\mu} +\frac{\kappa^2\ell(\ell +1)\mu{g^\prime}^2}{2e^2g^2} +\frac{\kappa^2\ell(2\ell+1)\mu a^2}{2e^2r^2} +V.
\ee
The use of \eqref{vinc} gives the interesting result 
\be
\begin{aligned}
\rho &= \frac{\kappa^2\ell\left(\ell+1\right)\mu}{2e^2g^2}\left(g^\prime \mp\sqrt{\frac{2\ell+1}{\ell+1}}\frac{ga}{r}\right)^2 \\
    &+\frac{1}{2\mu}\left(\sqrt{\frac{2\ell+1}{\ell}}\frac{a^\prime}{er} \pm\sqrt{2\mu V}\right)^2 \pm\frac1r\frac{dW}{dr},
\end{aligned}
\ee
where we have introduced
\be\label{aux}
	W(a,g) = -\frac{a}{e}\sqrt{\frac{2\left(2\ell+1\right)V(g)}{\ell\,\mu(g)}}.
\ee
This shows that the energy can be minimized when the potential is given by Eq.~\eqref{pot} and the first order equations \eqref{fo} are satisfied. The minimum energy is 
\be\label{ew}
E= 2\pi\left|W(a(\infty),g(\infty)) -W(a(0),g(0))\right|,
\ee
which only depends on the boundary values of $g$ and $a$, given by Eqs. \eqref{bcond}. We further notice that we have obtained the very same first order equations, although using a different procedure. We may then say that the solutions of the first order equations are stable under rescaling, and minimize the energy of the system as well.

Moreover, the energy density \eqref{rho} in this case can also be written in the form 
\be\label{dens}
\rho= \rho_G + \rho_E + \rho_B + \rho_V,
\ee
which describe the gradient, electric, magnetic and potential contributions in the energy density. They are given by
\bes
\bal
\rho_G &= \frac{\kappa^2\ell}{2e^2r^2}\left(3\ell+2\right)\mu(g)a^2 +\frac{\left(\ell+1\right)V(g)}{2\ell+1},\\
\rho_E &= \frac{\kappa^2\ell^2\mu(g)a^2}{2e^2r^2},\\
\rho_B &= \frac{\ell V(g)}{2\ell+1},\\
\rho_V &= V(g),
\eal
\ees
in which Eqs.~\eqref{fo} and \eqref{h} were used. From Eqs.~\eqref{lim} and \eqref{gori}, we see that each contribution above in the energy density is zero at $r=0$, for vortex solutions with any winding number.

Let us now provide an example for the model described by the Lagrange density that appears in Eq.~\eqref{model}. For convenience, we make the rescale
\be
\begin{aligned}
\vphi\to\frac{\kappa}{e}\vphi, \quad A_\alpha\to\frac{\kappa}{e}A_\alpha, \quad x_\alpha\to \frac{1}{\kappa}x_\alpha, \quad \LL\to\frac{\kappa^4}{e^2}\LL,
\end{aligned}
\ee
and also $v\to\kappa v/e$. This makes the fields and coordinates become dimensionless. We then proceed by taking $e=\kappa=1$ in the equations without loosing generality. We remark that the above rescale could be done right after Eq.~\eqref{model}. We are performing it here because we wanted to show how the parameters appear in the general equations.

\section{Some specific cases}
\label{sec3}

\begin{figure}[t!]
\centering
\includegraphics[width=4.2cm,trim={0.6cm 0.7cm 0 0},clip]{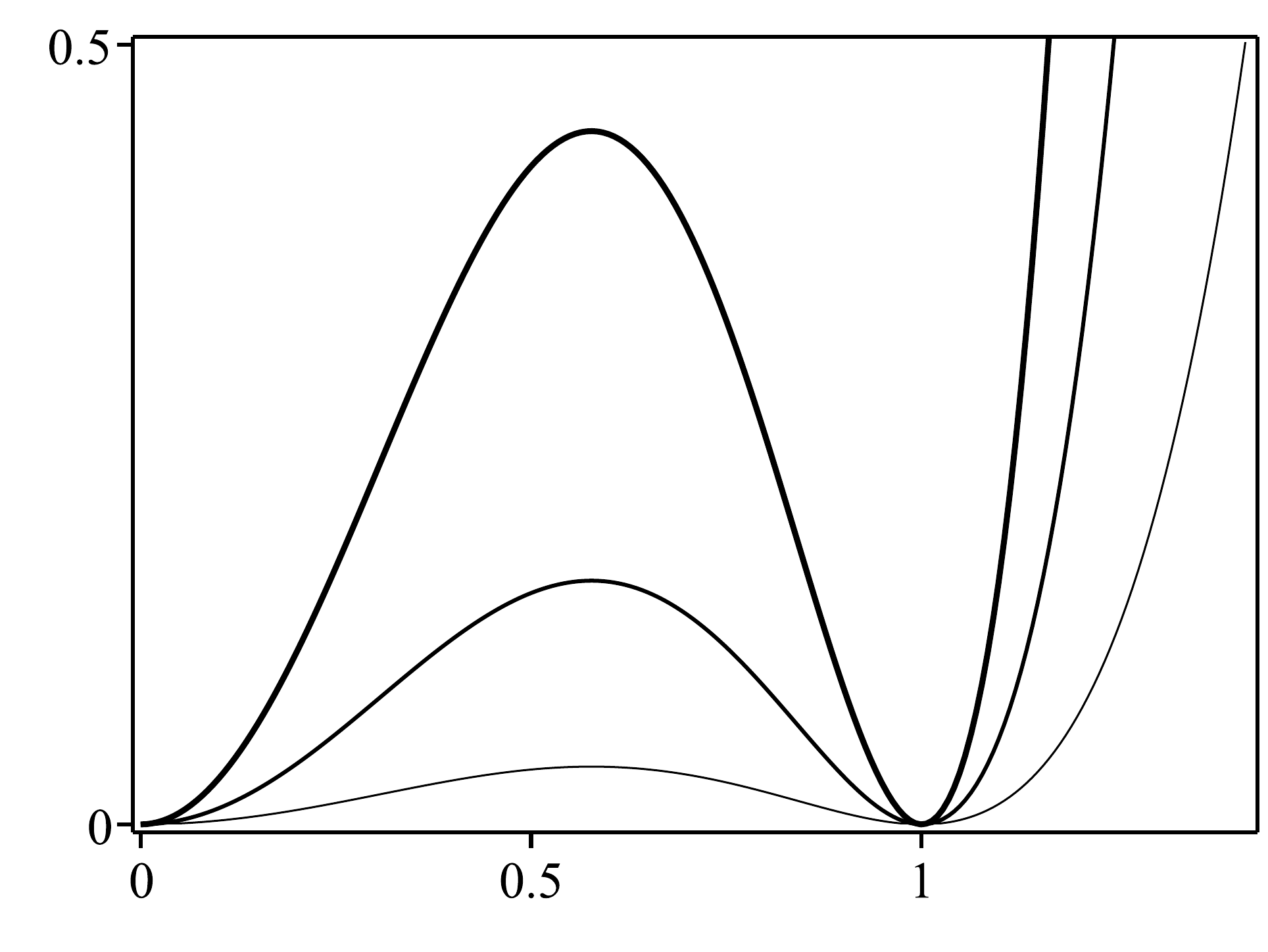}
\includegraphics[width=4.2cm,trim={0.6cm 0.7cm 0 0},clip]{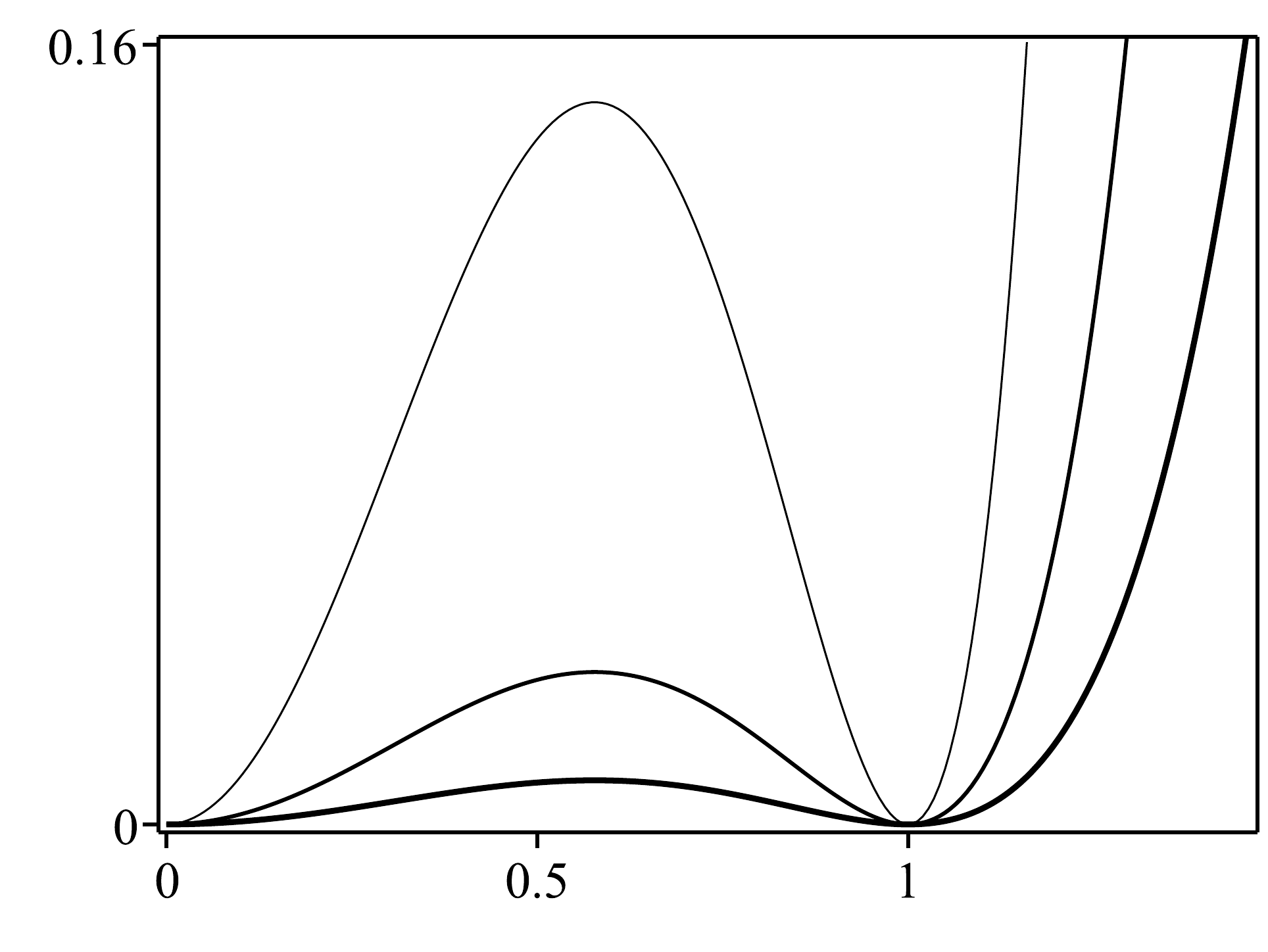}
\caption{The potential in Eq.~\eqref{pot1} for $v=c=1$. In the left panel one depicts the case with positive $\ell$, for $\ell=1,1.5$ and $2$, while the right panel displays the case with negative $\ell$, with $\ell=-2,-1.5$ and $-1.25$. In each panel, the thickness of the lines increases with increasing $\ell$.}
\label{fig1}
\end{figure}

To illustrate the procedure, let us first consider the magnetic permeability in the form
\be\label{mu1}
\mu(\vphia) = c\vphia^2.
\ee
with $c>0$. This choice leads to $M(\vphia)=c\ell(\ell+1)/2$, which does not depend on $\vphia$. As we have stated before, $\ell$ is a real parameter that must obey $\ell(\ell+1)>0$, which requires $\ell<-1$ or $\ell>0$. In this case, the potential in Eq.~\eqref{pot} becomes
\be\label{pot1}
V(\vphia) = \frac18c^3\ell^3\left(\ell+1\right)\vphia^2\left(v^2 -\vphia^2\right)^2,
\ee
where $v$ is a parameter involved in the symmetry breaking of the potential. This potential is shown in Fig.~\ref{fig1} for $v=c=1$ and some values of $\ell$.

The first order equations \eqref{fo} become
\bes\label{fo1}
\bal
g^\prime &= \pm\sqrt{\frac{2\ell+1}{\ell+1}}\frac{ag}{r},\\
-\frac{a^\prime}{r} &= \pm\frac12c^2\ell^2\sqrt{\frac{\ell+1}{2\ell+1}}g^2\left(v^2 -g^2\right).
\eal
\ees
and \eqref{h} takes the form
\be\label{h1}
h(r) = \pm\frac{c\ell}{2}\sqrt{\frac{\ell+1}{2\ell+1}}\left(v^2 -g^2(r)\right).
\ee
The boundary conditions \eqref{bcondh} for this solution are $h_0=\pm(c\ell v^2/2)\sqrt{(\ell+1)/(2\ell+1)}$ and $h_\infty=0$. In this procedure, Eqs.~\eqref{fo1} and \eqref{h1} determine the profile of the functions involved in the problem. A careful investigation shows that, asymptotically, the functions $g(r)$, $a(r)$, and $h(r)$ have to obey: $v-g(r>>0)\propto K_0\left(c\ell v^2r\right)$, $a(r>>0)\propto rK_1\left(c\ell v^2r\right)$ and $h(r>>0)\propto K_0\left(c\ell v^2r\right)$, where $K_\alpha(z)$ denotes modified Bessel function of the second kind.

\begin{figure}[t]
\centering
\includegraphics[width=4.2cm,trim={0.6cm 0.7cm 0 0},clip]{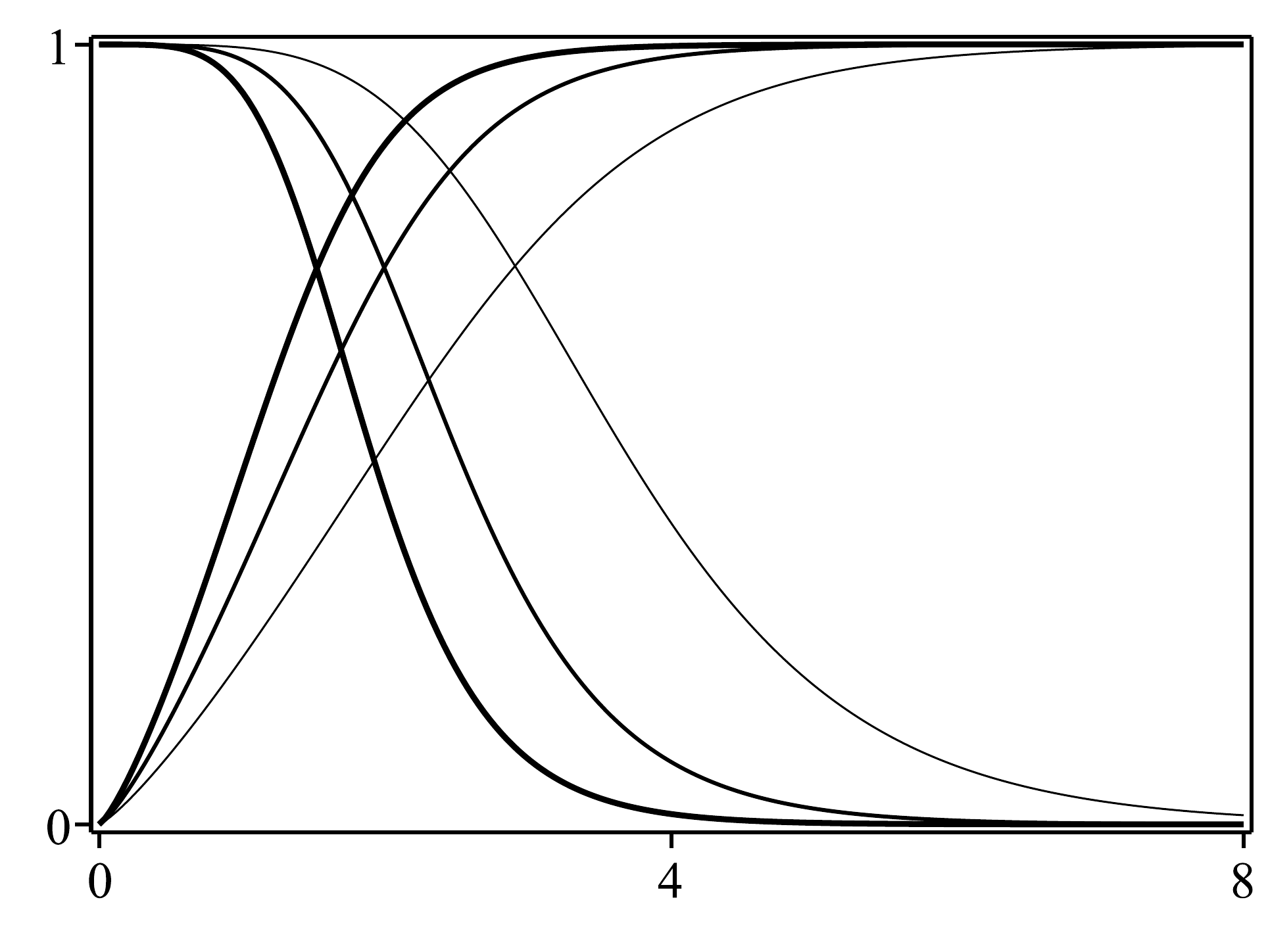}
\includegraphics[width=4.2cm,trim={0.6cm 0.7cm 0 0},clip]{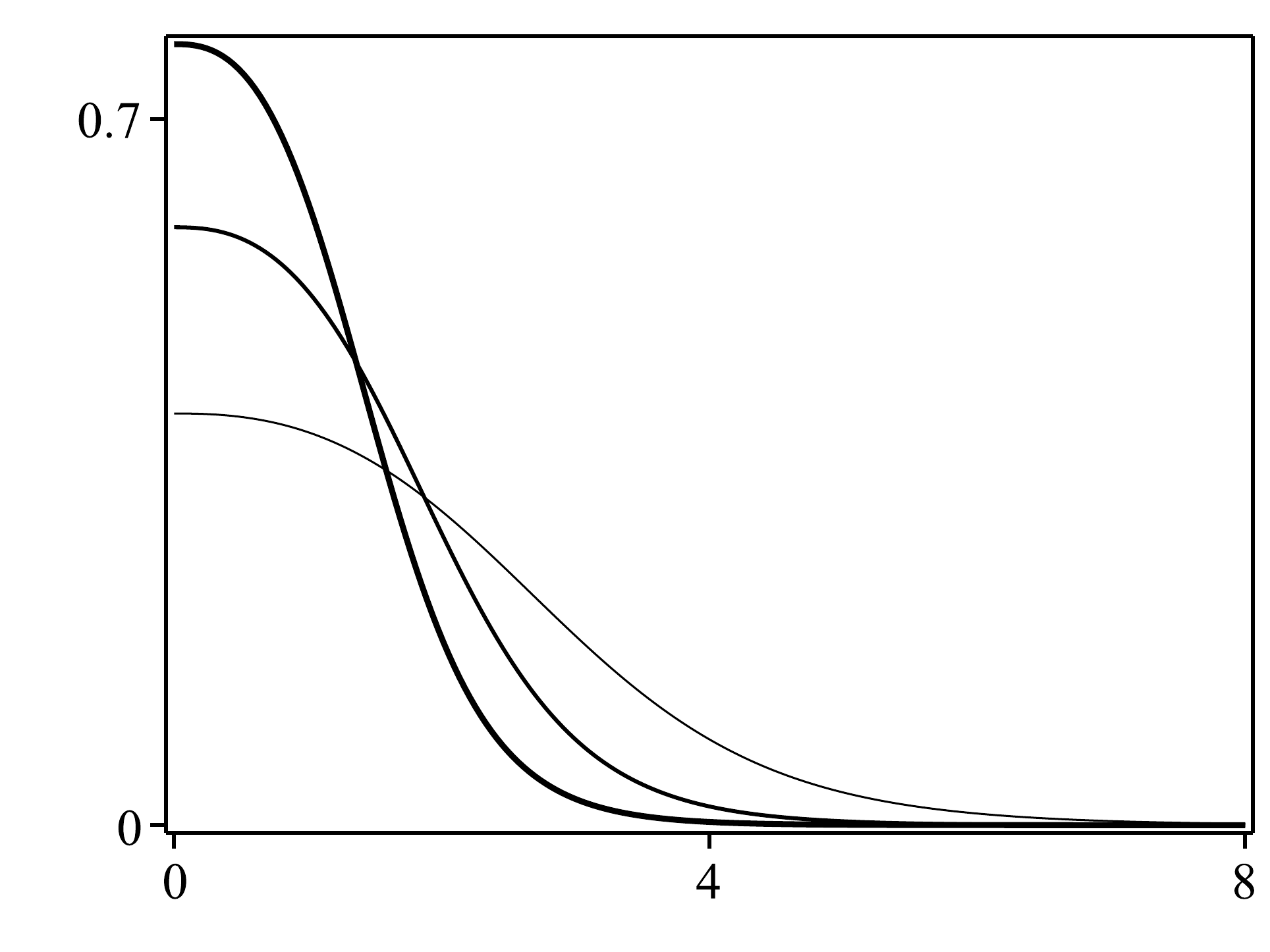}
\includegraphics[width=4.2cm,trim={0.6cm 0.7cm 0 0},clip]{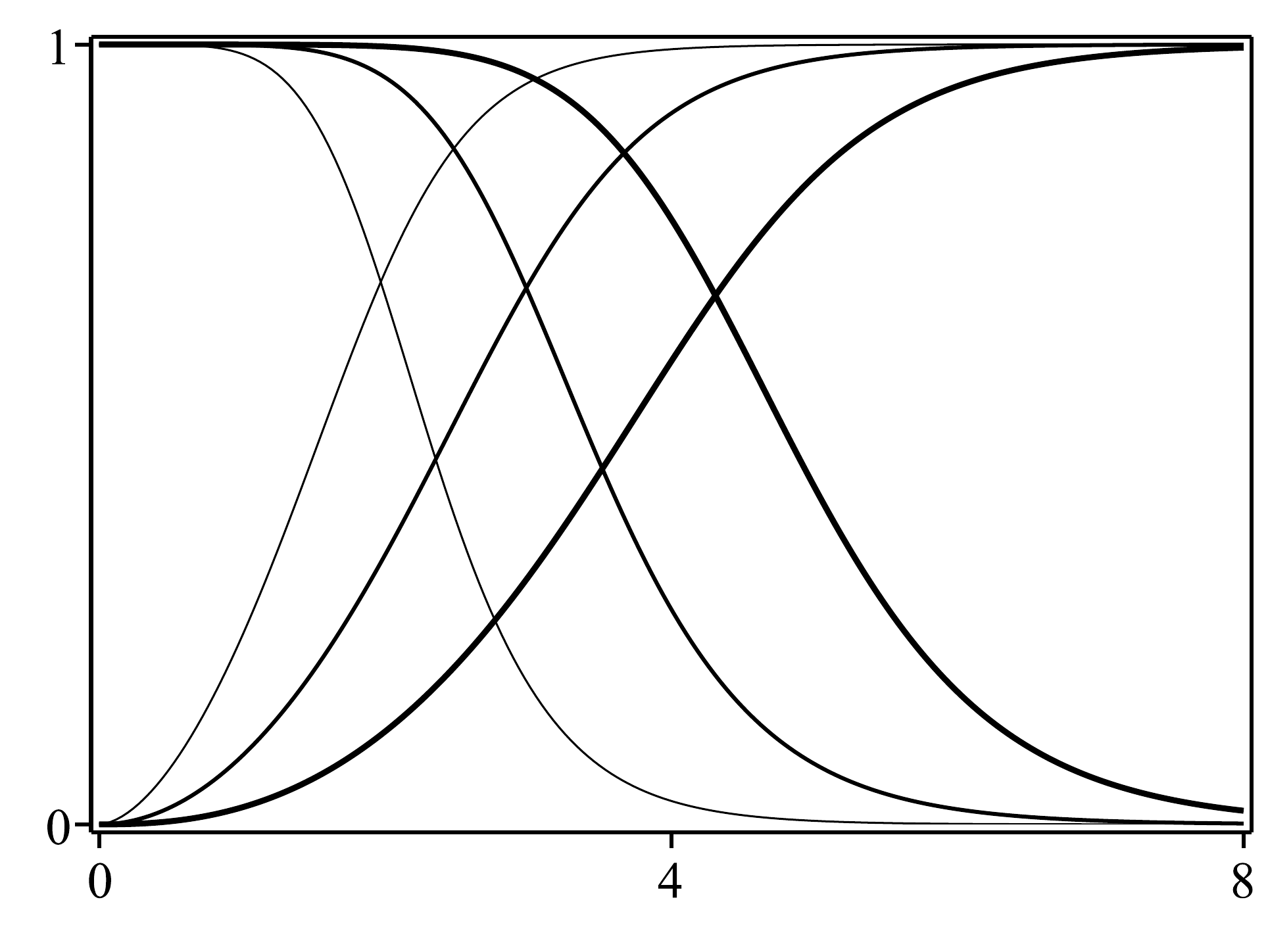}
\includegraphics[width=4.2cm,trim={0.6cm 0.7cm 0 0},clip]{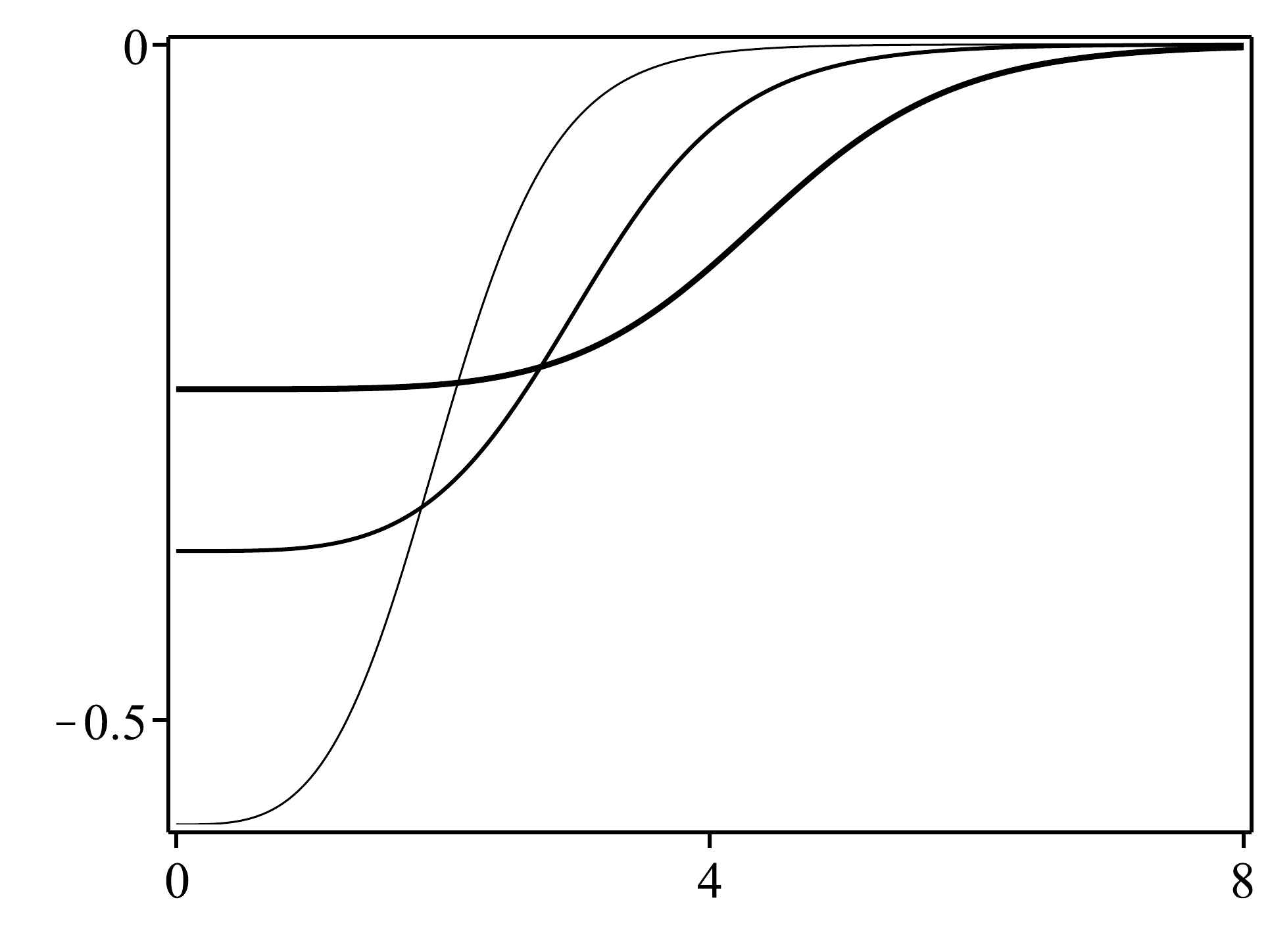}
\caption{In the left panels, one can see the solutions $g(r)$ (ascending lines) and $a(r)$ (descending lines) of Eq.~\eqref{fo1}. In the right panels, one has the profile of the function $h(r)$ in Eq.~\eqref{h1}, for $n=v=c=1$. The top panels depict the case with positive $\ell$, for $\ell=1,1.5$ and $2$, while the bottom panels are for negative $\ell$, with $\ell=-2,-1.5$ and $-1.25$. The thickness of the lines increases with $\ell$ in each panel.}
\label{fig2}
\end{figure}

\begin{figure}[bth!]
\centering
\includegraphics[width=4.2cm,trim={0.6cm 0.7cm 0 0},clip]{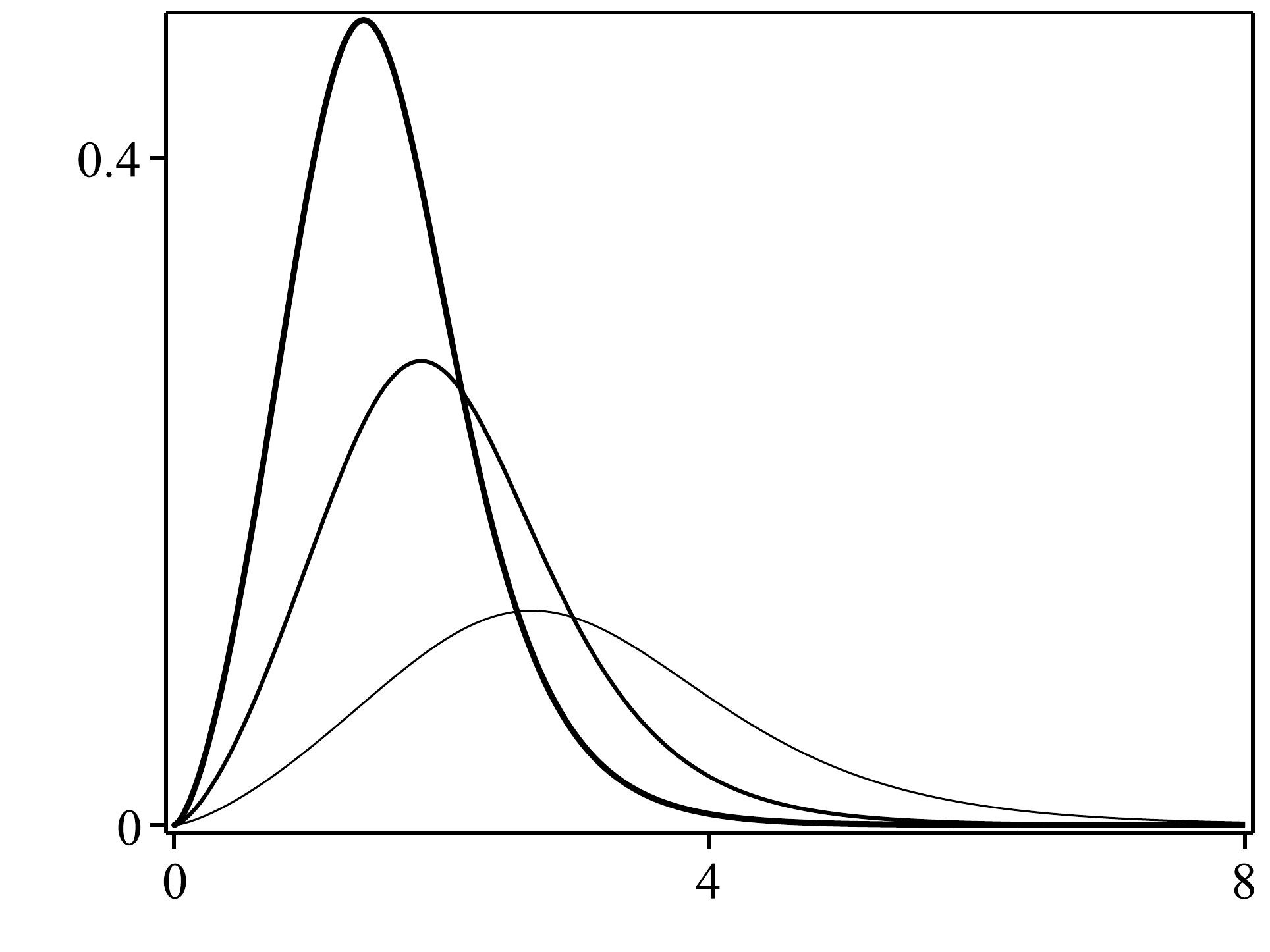}
\includegraphics[width=4.2cm,trim={0.6cm 0.7cm 0 0},clip]{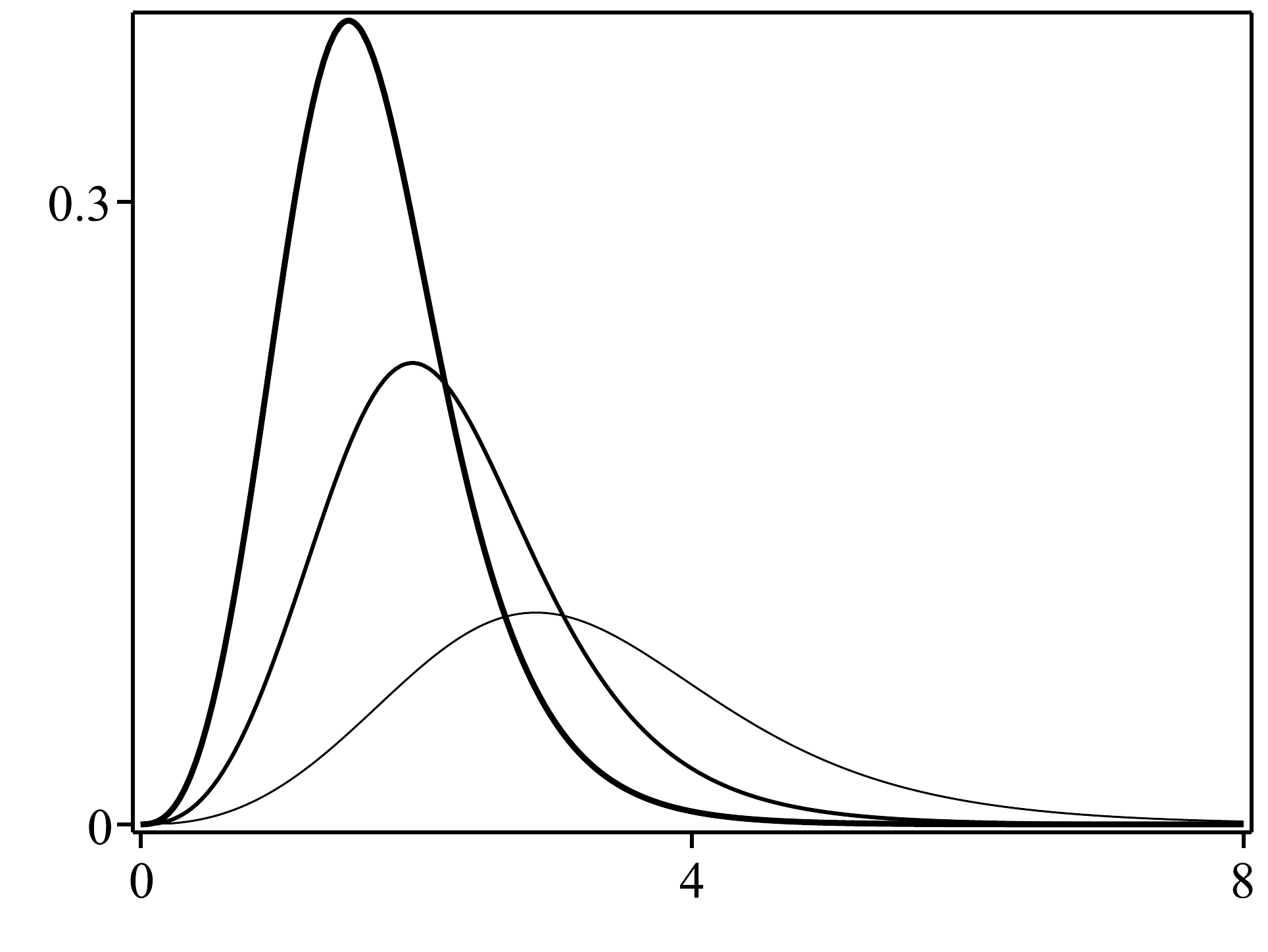}
\includegraphics[width=4.2cm,trim={0.6cm 0.7cm 0 0},clip]{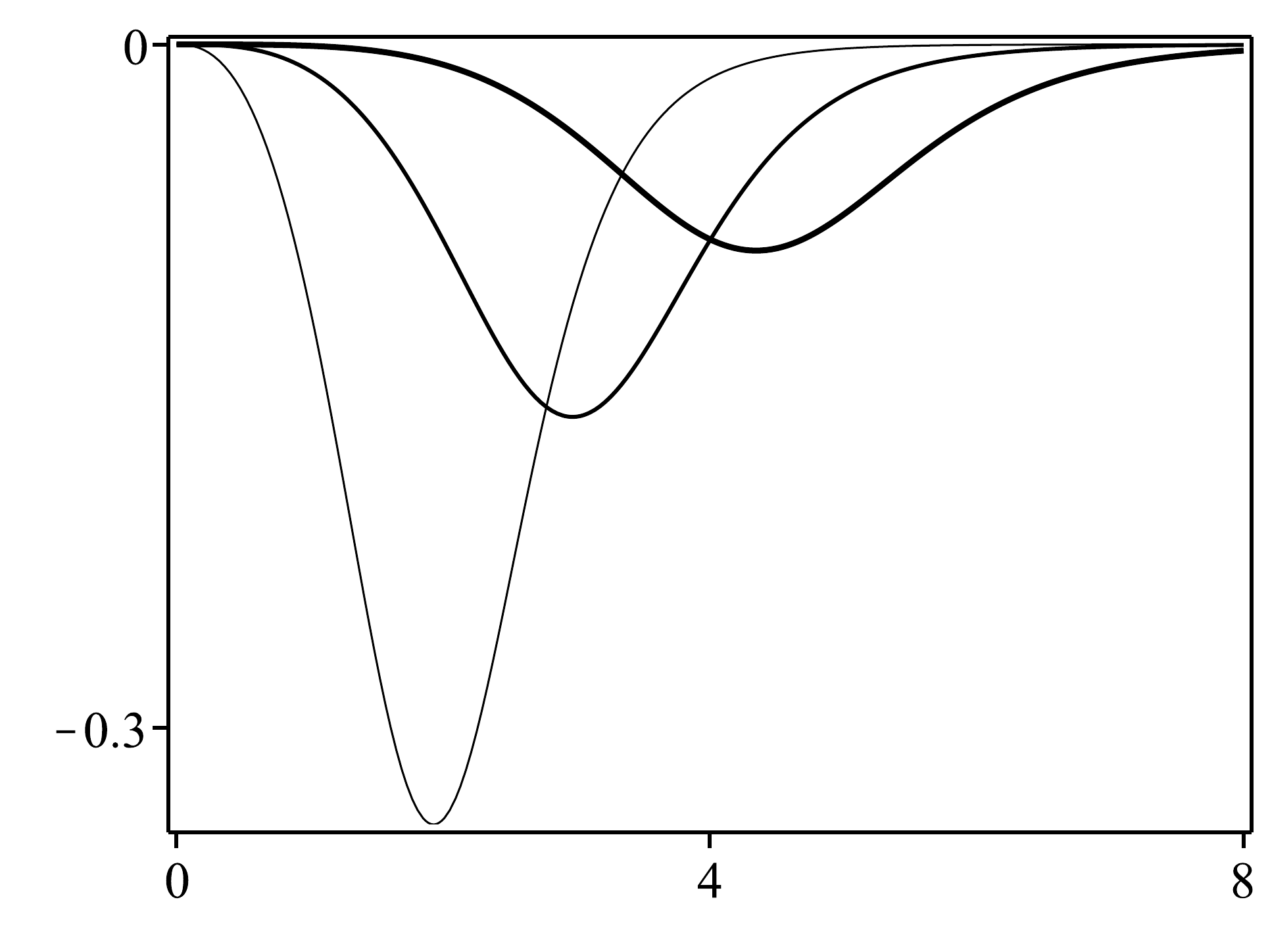}
\includegraphics[width=4.2cm,trim={0.6cm 0.7cm 0 0},clip]{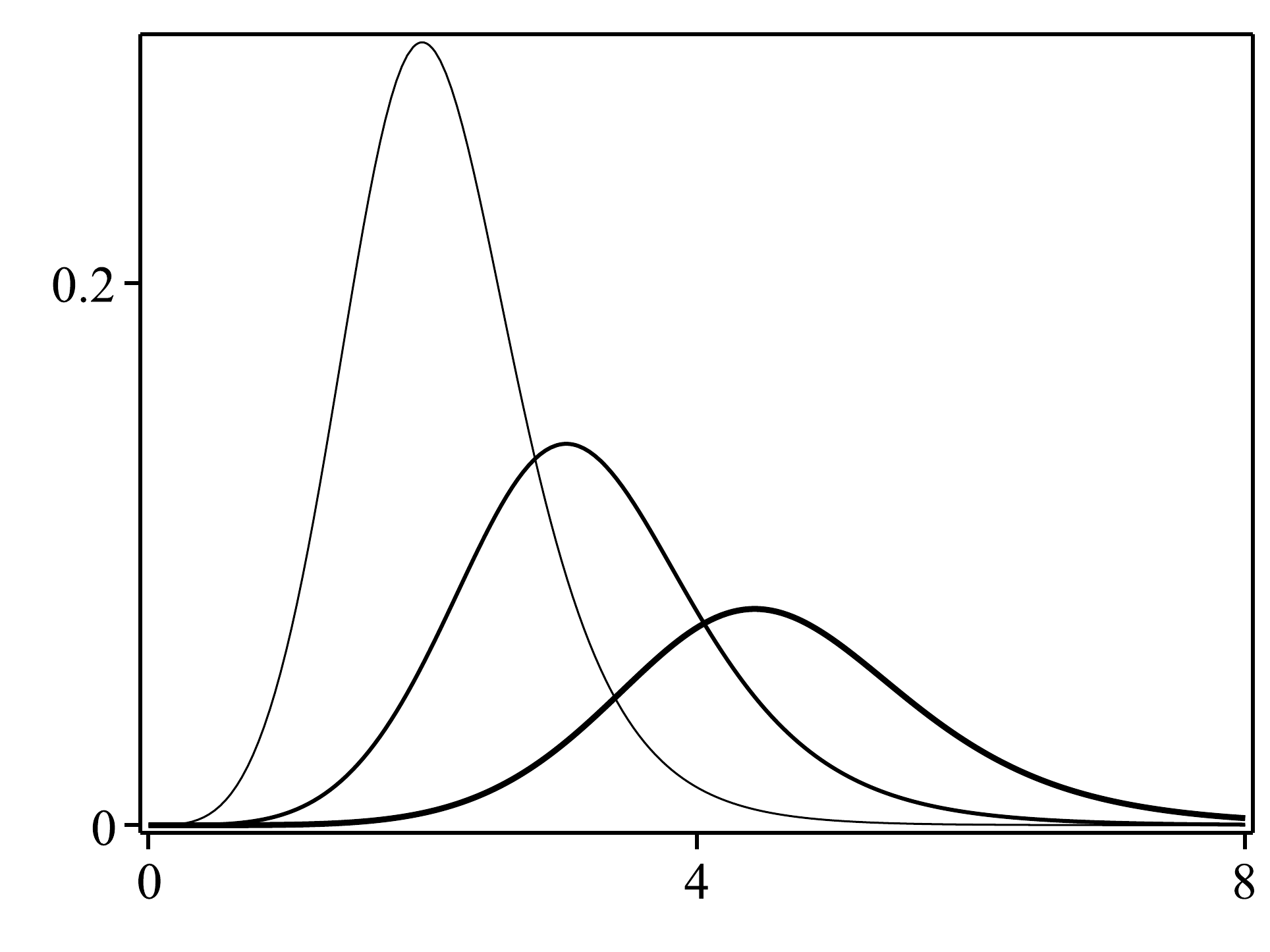}
\caption{The radial electric field (left) and the magnetic field (right) in Eq.~\eqref{fieldsEB} associated to Eqs.~\eqref{fo1} and \eqref{h1}, for $n=v=c=1$. The top panels depict the case with positive $\ell$, for $\ell=1,1.5$ and $2$, while the bottom panels are for negative $\ell$, with $\ell=-2,-1.5$ and $-1.25$. The thickness of the lines increases with $\ell$ in each panel.}
\label{fig3}
\end{figure}

Moreover, near the origin the function $g(r)$ behaves as in Eq.~\eqref{gori}, and $a(r)$ and $h(r)$ obey
\bes
\bal
\label{aori}n-a(r\approx0)&\propto r^{2\left(1+|n|\sqrt{(2\ell+1)/(\ell+1)}\right)},\\
h_0-h(r\approx0)&\propto r^{2|n|\sqrt{(2\ell+1)/(\ell+1)}}.
\eal
\ees
The electric and magnetic fields can be calculated from Eq.~\eqref{fieldsEB}. The behavior of these quantities near the origin is given by
\bes
\bal
E_r(r\approx 0)&\propto r^{2|n|\sqrt{(2\ell+1)/(\ell+1)}-1},\\
B(r\approx 0)&\propto r^{2|n|\sqrt{(2\ell+1)/(\ell+1)}}.
\eal
\ees
In Figs.~\ref{fig2} and \ref{fig3}, we display the profile of solutions and their respective physical quantities, the electric and magnetic fields, for $n=v=c=1$ and some values of $\ell$. We notice that the electric field changes sense as one changes from positive to negative $l$; this is in agreement with the change of sign of the electric charge, as we commented on in the paragraph below Eq. \eqref{bh}, and is another feature, which is not present is the Maxwell-Chern-Simons model in the presence of the neutral field. 

\begin{figure}[t!]
\centering
\includegraphics[width=4.2cm,trim={0.6cm 0.7cm 0 0},clip]{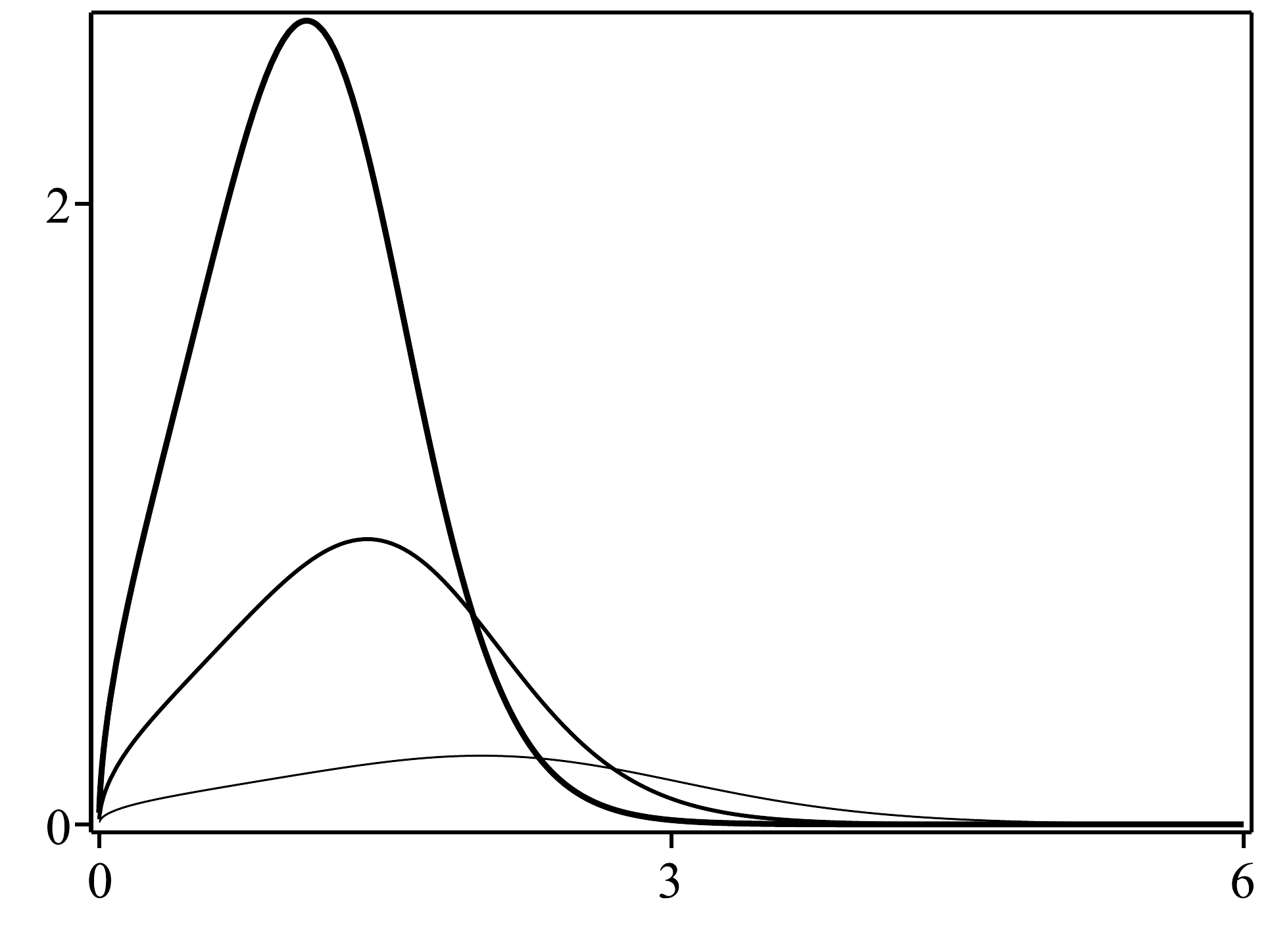}
\includegraphics[width=4.2cm,trim={0.6cm 0.7cm 0 0},clip]{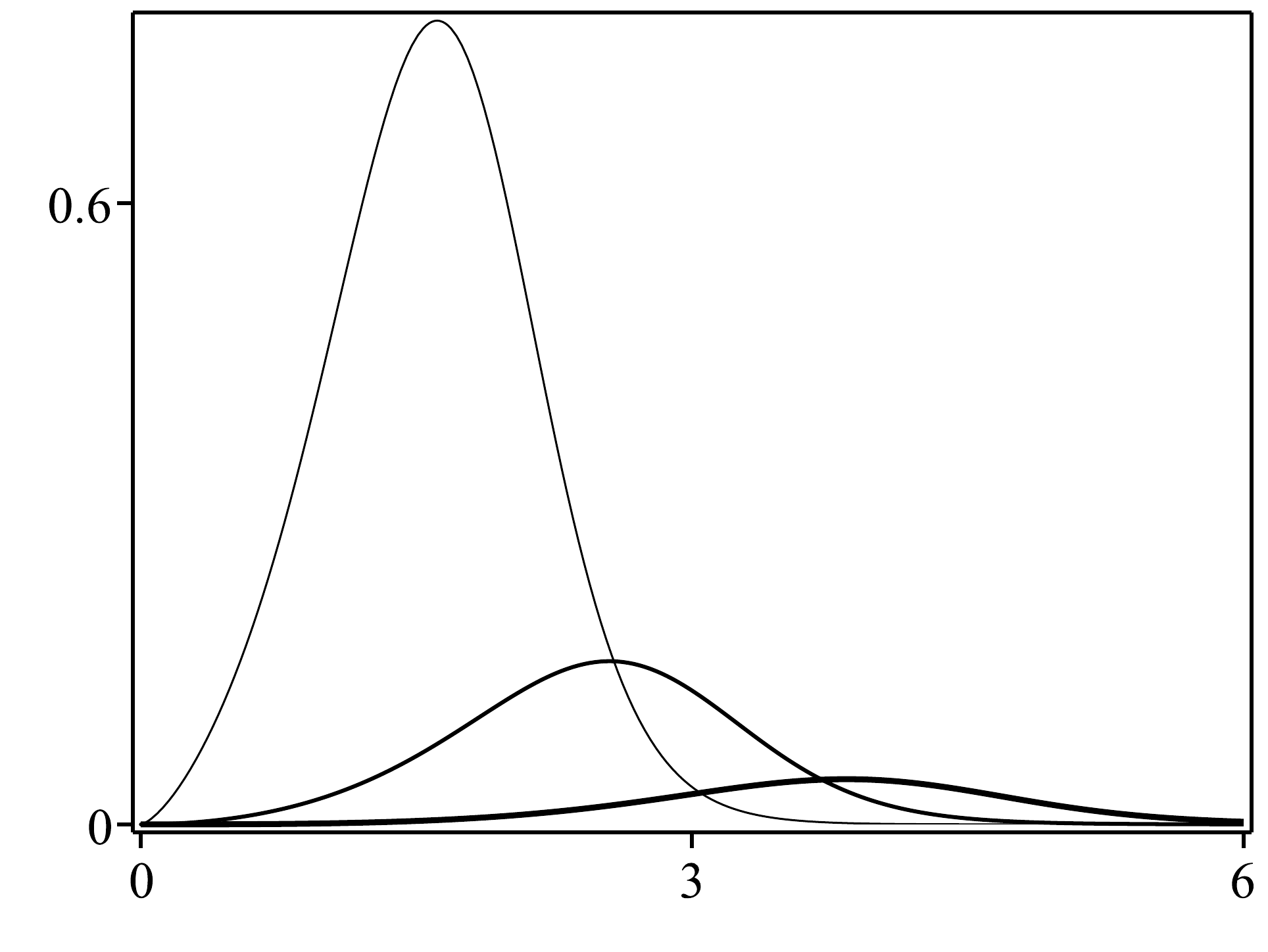}
\caption{The energy density in Eq.~\eqref{rho1} for $n=v=c=1$ and $\ell=1,1.5$ and $2$ (left) and $\ell=-2,-1.5$ and $-1.25$ (right). The thickness of the lines increases with $\ell$.}
\label{fig4}
\end{figure}

\begin{figure}[bth!]
\centering
\includegraphics[width=4.2cm,trim={0.6cm 0.7cm 0 0},clip]{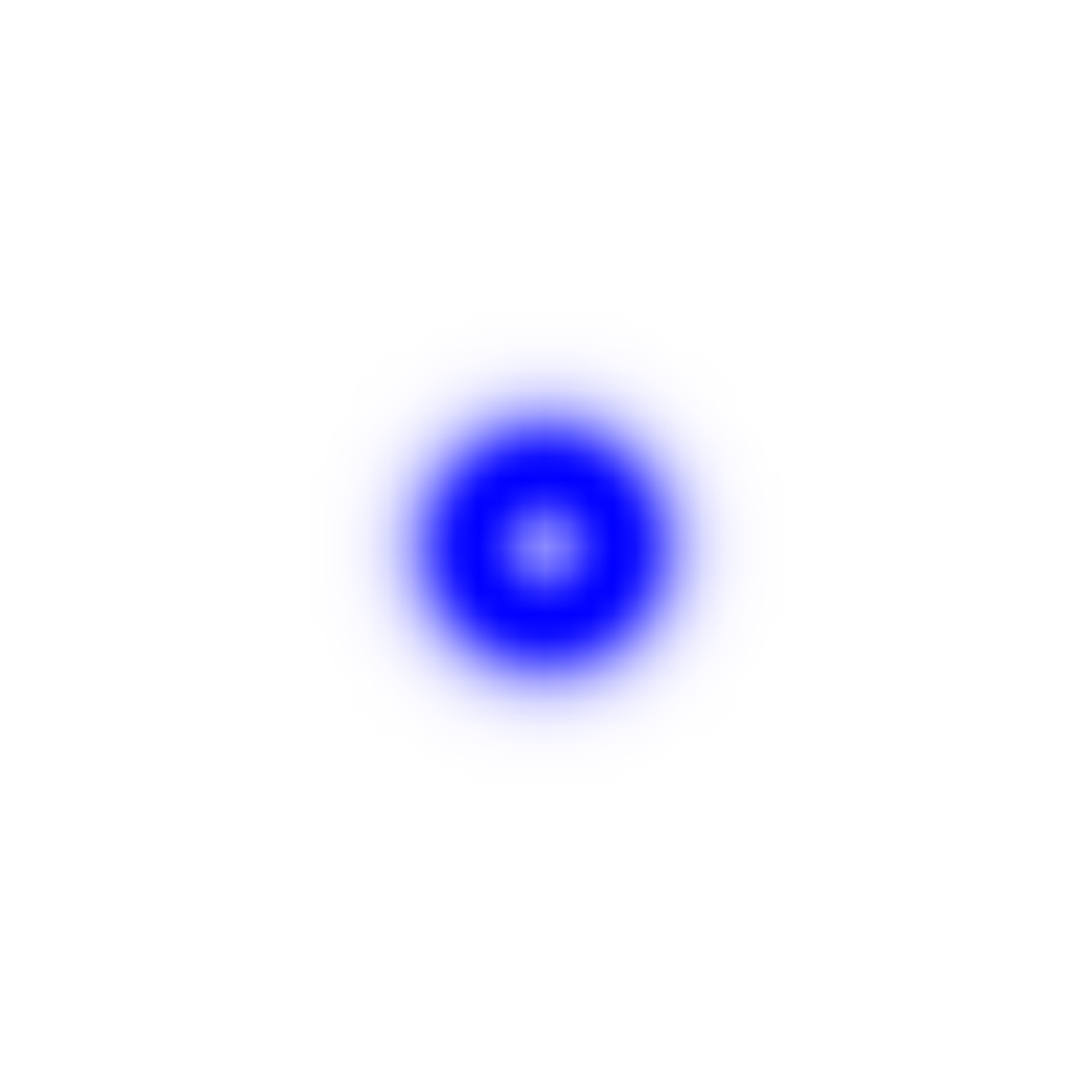}
\includegraphics[width=4.2cm,trim={0.6cm 0.7cm 0 0},clip]{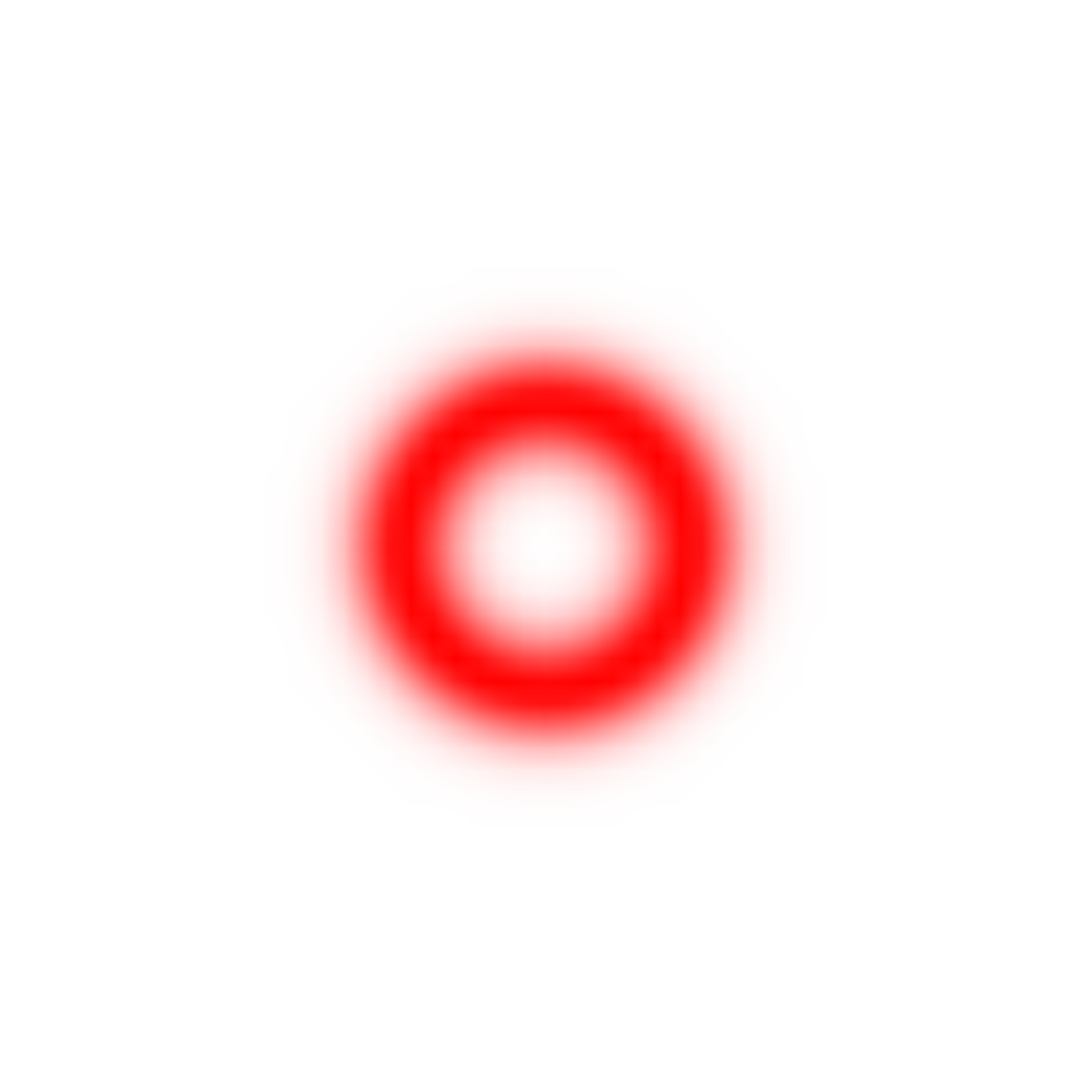}
\caption{The energy density in Eq.~\eqref{rho1} for $n=v=c=1$ and $\ell=1$ (blue) and $\ell=-1.25$ (red) in the plane. The intensity of the colors increases with the increasing values of the energy densities.}
\label{fig5}
\end{figure}

The energy density in Eq.~\eqref{dens} becomes
\be\label{rho1}
\rho = c\ell\left(2\ell+1\right)g^2\left(\frac{a^2}{r^2}+\frac{c^2\ell^2\left(\ell+1\right)}{4\left(2\ell+1\right)}\left(v^2 -g^2\right)^2\right).
\ee
The solutions are numerical, but the behavior of the energy density near $r=0$ can be found with the approximate form of $a(r)$ given by Eq.~\eqref{aori}, and $g(r)$ by Eq.~\eqref{gori}. It has the form:
\be
\rho(r\approx0) \propto r^{2\left(|n|\sqrt{(2\ell+1)/(\ell+1)}-1\right)}.
\ee
The general behavior of the energy density \eqref{rho1} can be seen in Fig.~\ref{fig4} for $n=v=c=1$ and some values of $\ell$. Notice that there is a hole around the origin as expected from the above expression. To highlight this feature, we depict the energy density in the plane in Fig.~\ref{fig5}. The auxiliary function in Eq.~\eqref{aux} becomes
\be
W(g,a) = -\frac{c}{2}\sqrt{\ell^2\left(\ell+1\right)\left(2\ell+1\right)}\,a\left(v^2 -g^2\right).
\ee
So, the energy \eqref{ew} is
\be
E = \pi cv^2|n|\sqrt{\ell^2\left(\ell+1\right)\left(2\ell+1\right)}.
\ee


We emphasize that the Maxwell-Chern-Simons model in Eq.~\eqref{model} does not include the neutral field to obtain vortices described by first order equations as it was done in Refs.~\cite{nmcs,bazeiamcs,menezesmcs,godvortex}. We have shown that rotationally symmetric solutions stable under rescale are obtained from the stressless condition, $T_{ij}=0$. Also, for $M(\vphia)$ and $\mu(\vphia)$ obeying Eq.~\eqref{M} and taking Eq.~\eqref{hprime}, the equations of motion \eqref{scalareom} and \eqref{mcseom} can be reduced to first order, in the form \eqref{fo}. This method requires the potential to have the form \eqref{pot}.

The first order equation \eqref{fog} presents a factor controlled by $\ell$ that modifies the behavior of the function $g(r)$, associated to the scalar field, at the origin. As it was shown in Ref.~\cite{godvortex}, in Maxwell, Chern-Simons and Maxwell-Chern-Simons models with neutral field, the equation for $g(r)$ is $g'=ag/r$. So, our procedure introduces a modification in this equation that takes an important role in the electric and magnetic fields, and also in the energy density.

\begin{figure}[bth!]
\centering
\includegraphics[width=4.2cm,trim={0.6cm 0.7cm 0 0},clip]{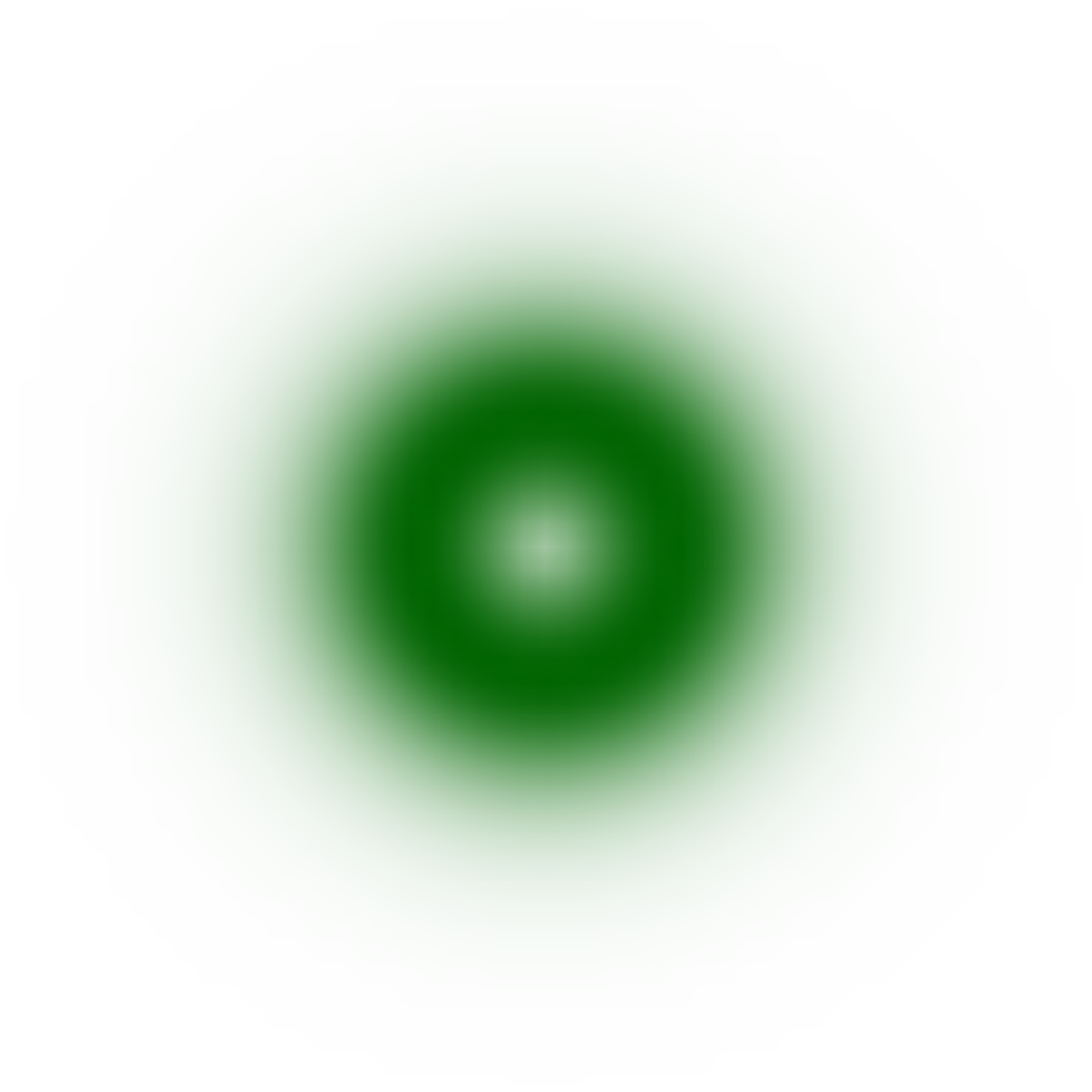}
\includegraphics[width=4.2cm,trim={0.6cm 0.7cm 0 0},clip]{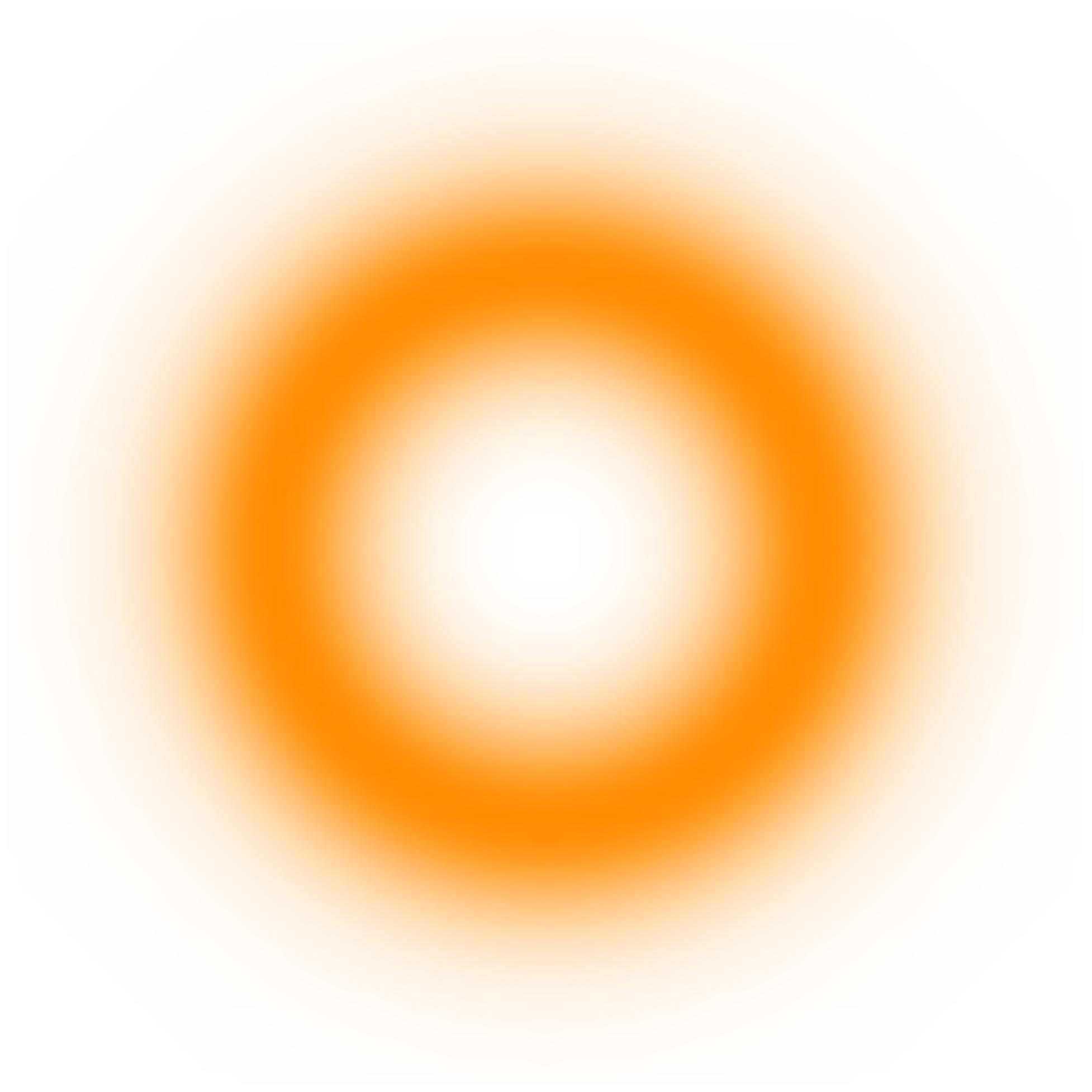}
\caption{The energy density in the new model, for $n=v=c=1$ and $\ell=1$ (green) and $\ell=-1.25$ (orange) in the plane. The intensity of the colors increases as in Fig. \ref{fig5}.}
\label{fig6}
\end{figure}

Up to here, we have studied a model with the magnetic permeability described by Eq.~\eqref{mu1}, which leads to a constant $M(\vphia)$. The aforementioned modification in the first order equations leads to the presence of a hole with size controlled by $\ell$ around the origin in all the contributions of the energy density in Eq.~\eqref{dens} for this situation, with unit vorticity. It is of interest to remark that, even though we have investigated a model with a constant $M(\vphia)$, other possibilities for the generalized magnetic permeability can also be studied. One must, however, take into account that the constraints \eqref{M} and \eqref{vinc} have to be satisfied. For instance, we have considered another case, with $\mu(\vphia) = c\vphia^2|v^2-\vphia^2|$ and $M(\vphia)=c\ell(\ell+1)|v^2-\vphia^2|/2$: we have checked that it leads to a model which supports finite energy solutions compatible with the boundary conditions \eqref{bcond} and the equations of motion. The procedure follows as before, and to show how the energy density behaves in this case, we display it in Fig. \ref{fig6}. The profile is qualitatively similar to the one depicted in Fig. \ref{fig5}, but the green and orange regions in Fig. \ref{fig6} are almost twice as big as the blue and red ones that appeared in Fig. \ref{fig5}, depicted at the same scale.\\

\section{Ending comments}
\label{end}

We think that the procedure introduced in this paper may fosters new studies in the subject. In particular, we are now searching for other models, to see how one can modify the asymptotic behavior of these novel vortex configurations. Moreover, since in general the presence of first order equations suggests the existence of supersymmetric extension, it appears appropriate to consider this issue with the model considered in this work.
The presence of supersymmetry is of current interest in high energy physics \cite{susy}, and the possibility to use it in connection with the model studied in this work may be related to the previous investigation \cite{ba}, which explore supersymmetric extension of model described by field with generalized kinetic term, to the work \cite{adam}, in which the authors construct a method to supersymmetrize higher kinetic terms to apply it to the baby Skyrme model, and also to \cite{queiruga}, in which one investigates several aspects of supersymmetric three-dimensional higher-derivative field theories. We can also follow the lines of \cite{research} and deal with the possibility to enhance the gauge symmetry to accommodate additional fields, responsible to generate localized structures having the form of multilayered vortices. Another motivation to include the generalized magnetic permeability $\mu(\vphia)$ and $M(\vphia)$, is inspired by the unconventional superconductivity recently observed in twisted bilayer \cite{nature} and trilayer graphene \cite{T1,T2}. These results show that the addition of extra (appropriately twisted) layers in the standard (monolayer) graphene induce the appearance of unconventional properties in the physical system under investigation. In the generalized MCS system here studied, the presence of $\mu(\vphia)$ and $M(\vphia)$ allows the construction of first order framework, capable of describing vortices with unconventional features.

\acknowledgements{This work is supported by the Brazilian agencies Coorden\c{c}\~ao de Aperfei\c{c}oamento de Pessoal de N\'ivel Superior (CAPES), grant No 88887.485504/2020-00 (MAL), Conselho Nacional de Desenvolvimento Cient\'ifico e Tecnol\'ogico (CNPq), grants Nos. 140490/2018-3 (IA), 404913/2018-0 (DB), 303469/2019-6 (DB) and 306504/2018-9 (RM), Paraiba State Research Foundation (FAPESQ-PB) grants Nos. 0003/2019 (RM) and 0015/2019 (DB, MAL and MAM) and by Federal University of Para\'iba (PROPESQ/PRPG/UFPB) project code PII13363-2020.}


\begin{thebibliography}{99}
\bibitem{NO} H.B. Nielsen and P. Olesen, 
Nucl. Phys. B {\bf61}, 45 (1973).
\bibitem{coreanos}J. Hong, Y. Kim and P.Y. Pac, 
Phys. Rev. Lett. {\bf64}, 2230 (1990).
\bibitem{jackiw}R. Jackiw and E.J. Weinberg, 
Phys. Rev. Lett. {\bf 64}, 2234 (1990).
\bibitem{paulkhare}S.K. Paul and A. Khare, 
Phys. Lett. B {\bf174}, 420 (1986). [\textit{Erratum ibid} {\bf177}, 453 (1986)].
\bibitem{jacobs} L. Jacobs, A. Khare, C.N. Kumar, S.K. Paul, 
Int. J. Mod. Phys. A {\bf6}, 3441 (1991).
\bibitem{derrick} G.H. Derrick, 
J. Math. Phys. \textbf{5}, 1252 (1964).
\bibitem{bogopaper} E.B. Bogomol'nyi, 
Sov. J. Nucl. Phys {\bf24}, 449 (1976).
\bibitem{schaposnik} H.J. de Vega and F.A. Schaposnik, Phys. Rev. D \textbf{14}, 1100 (1976).
\bibitem{godvortex} D. Bazeia, L. Losano, M.A. Marques, R. Menezes and I. Zafalan, 
Nucl. Phys. B {\bf 934}, 212 (2018).
\bibitem{nmcs}C. Lee, K. Lee, and H. Min, 
Phys. Lett. B {\bf252}, 79 (1990).
\bibitem{bazeiamcs} D. Bazeia, 
Phys. Rev. D {\bf43}, 4074 (1991).
\bibitem{menezesmcs}  D. Bazeia, R. Casana, E. da Hora, R. Menezes, 
Phys. Rev. D {\bf85}, 125028 (2012).
\bibitem{torres} M. Torres, 
Phys. Rev. D {\bf 46}, R2295 (1992).
\bibitem{ghoshplb} P.K. Ghosh, 
Phys. Lett. B {\bf 326}, 264 (1994),
\bibitem{ghosh} P.K. Ghosh, 
Phys. Rev. D {\bf 49}, 5458 (1994).
\bibitem{nminimo} I. Andrade, D. Bazeia, M.A. Marques, R. Menezes, 
Phys. Rev. D {\bf 102}, 045018 (2020).
\bibitem{susy}M. Dine, {\it Supersymmetry and String Theory}, Cambridge University Press, 2015.
\bibitem{ba}D. Bazeia, R. Menezes, and A.Yu. Petrov, Phys. Lett. B {\bf683}, 335  (2010).
\bibitem{adam} C. Adam, J.M. Queiruga, J. Sanchez-Guillen, and 
A. Wereszczynski, Phys. Rev. D {\bf84}, 025008 (2011).
\bibitem{queiruga}J.M. Queiruga, Phys. Rev. D {\bf95}, 125001 (2017).
\bibitem{research}D. Bazeia, M.A. Liao, M.A. Marques, and R. Menezes, Phys. Rev. Research {\bf1}, 033053 (2019). 
\bibitem{nature}Y. Cao, et al. 
Nature {\bf556}, 43 (2018).
\bibitem{T1}J.M. Park, et al. Nature {590}, 249 (2021).
\bibitem{T2}Z. Hall et al. Science {\bf371}, 1133 (2021).
\end{thebibliography}
\end{document}